\documentclass[aps,prd,preprint,12pt,superscriptaddress,nofootinbib,a4paper]{revtex4-1}

\usepackage[utf8]{inputenc}
\usepackage{amsmath,amssymb}
\usepackage[separate-uncertainty=true]{siunitx}
\usepackage{graphicx}
\usepackage[usenames,dvipsnames]{xcolor}
\usepackage[caption=false]{subfig}
\usepackage{hyperref}
\usepackage[capitalize]{cleveref}
\usepackage{booktabs}
\usepackage{tabularx}
\usepackage{xspace}
\usepackage{color}
\usepackage[normalem]{ulem}
\usepackage{slashed}
\usepackage{bbold}
\usepackage{wasysym}
\usepackage{graphicx}
\usepackage{mathrsfs} 

\allowdisplaybreaks

\graphicspath{{graphics/}}

\newcommand{\eqn}{equation}
\newcommand{\lb}{\left(}
\newcommand{\rb}{\right)}

\newcommand{\al}{\alpha}

\newcommand{\GeV}{{\ensuremath\rm GeV}}

\newcommand{\pb}{{\ensuremath\rm pb}}
\newcommand{\fb}{{\ensuremath\rm fb}}
\newcommand{\ab}{{\ensuremath\rm ab}}

\DeclareSIUnit{\pb}{pb}
\DeclareSIUnit{\fb}{fb}
\AtBeginDocument{
\heavyrulewidth=.08em
\lightrulewidth=.05em
\cmidrulewidth=.03em
\belowrulesep=.65ex
\belowbottomsep=0pt
\aboverulesep=.4ex
\abovetopsep=0pt
\cmidrulesep=\doublerulesep
\cmidrulekern=.5em
\defaultaddspace=.5em

\newcolumntype{C}{>{\centering\arraybackslash}X}
\newcolumntype{b}{C}
\newcolumntype{s}{>{\hsize=.6\hsize}C}
\newcolumntype{R}{>{\raggedleft\arraybackslash}X}
}

\begin{document}
\date{\today}
\rightline{RBI-ThPhys-2023-23, CERN-TH-2023-149}
\title{{\Large A short overview on low mass scalars\\ at future lepton colliders \\ -- LCWS23 proceedings --}}
\author{Tania Robens}
\affiliation{Ruder Boskovic Institute, Bijenicka cesta 54, 10000 Zagreb, Croatia\\ and \\
Theoretical Physics Department, CERN, 1211 Geneva 23, Switzerland}

\renewcommand{\abstractname}{\texorpdfstring{\vspace{0.5cm}}{} Abstract}

\begin{abstract}
    \vspace{0.5cm}
I give a short summary on scenarios with new physics scalars that could be investigated at future $e^+e^-$ colliders. I concentrate on cases where at least one of the additional scalar has a mass below 125 GeV, and discuss both models where this could be realized, as well as studies which focus on such scenarios. This work is based on \cite{Robens:2022zgk}, and partial results were also presented in \cite{Robens:2022uis}.
  
\end{abstract}

\maketitle

\section{Introduction}
In the European Strategy report \cite{EuropeanStrategyforParticlePhysicsPreparatoryGroup:2019qin,CERN-ESU-015}, Higgs factories were identified as one of the high priority projects after the HL-LHC. At such machines, the properties of the Higgs particles should be measurable to utmost precision. Furthermore, new physics scalar states could also be produced in the mass range up to $\sim\,160\,\GeV$ depending on the collider process.

At LEP, a large variety of new physics models have been investigated, with a concise summary given in \cite{OPAL:2002ifx,ALEPH:2006tnd}. However, these searches, as well as more recent bounds from direct searches for light scalars at the LHC, still leave room for new light scalar states in many new physics models. In this short proceeding, I summarize the current status of such searches and present new physics scenarios that still allow for such states, taking all current constraints into account. The work presented here is based on \cite{Robens:2022zgk}, and some of the results have additionally been presented in \cite{Robens:2022uis}.

\section{Processes at Higgs factories}

\begin{center}
\begin{figure}[htb!]
\begin{center}
\includegraphics[width=0.55\textwidth]{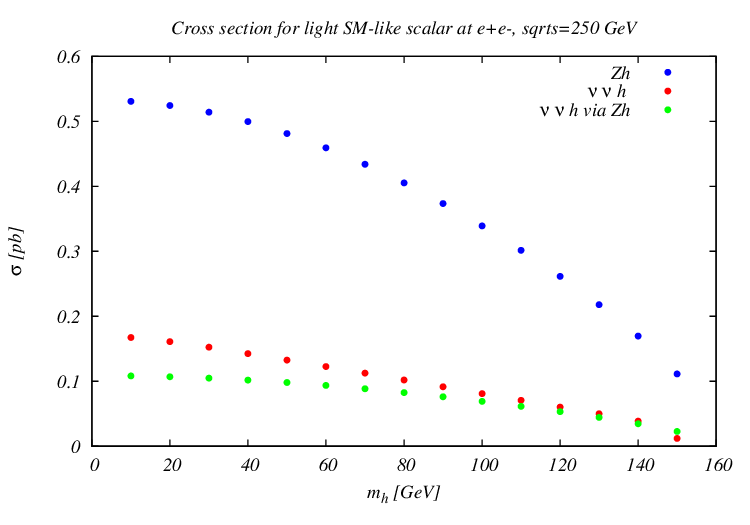}
\caption{\label{fig:prod250} Leading-order production cross sections for $Z\,h$ and $h\,\nu_\ell\,\bar{\nu}_\ell$ production at an $e^+\,e^-$ collider with a com energy of 250 \GeV~  using Madgraph5 for an SM-like scalar $h$. Shown is also the contribution of $Z\,h$ to $\nu_\ell\,\bar{\nu}_\ell\,h$ using a factorized approach for the Z decay. Update of plot in \cite{Robens:2022zgk}, first presented in \cite{Robens:2022uis}.}
\end{center}
\end{figure}
\end{center}

At the center-of-mass (com) energies of Higgs factories, Higgs strahlung is the dominant production mode \cite{Abramowicz:2016zbo}. Leading-order predictions for $Zh$ production at $e^+e^-$ colliders for low mass scalars which are Standard Model (SM)-like, using Madgraph5 \cite{Alwall:2011uj}, are shown in figure \ref{fig:prod250} for a center-of-mass energy of 250 \GeV.  The VBF-type production of $e^+e^-\,\rightarrow\,h\,\nu_\ell\,\bar{\nu}_\ell$ contains contributions from $Z\,h$ production, where $Z\,\rightarrow\,\nu_\ell\,\bar{\nu}_\ell$. While for lower masses VBF production still plays a role, for higher scalar masses the dominant contribution stems from $Z\,h$ production.

\section{Projections for additional searches}
\begin{center}
\begin{figure}[htb!]
\begin{center}
\includegraphics[width=0.4\textwidth, angle=-90]{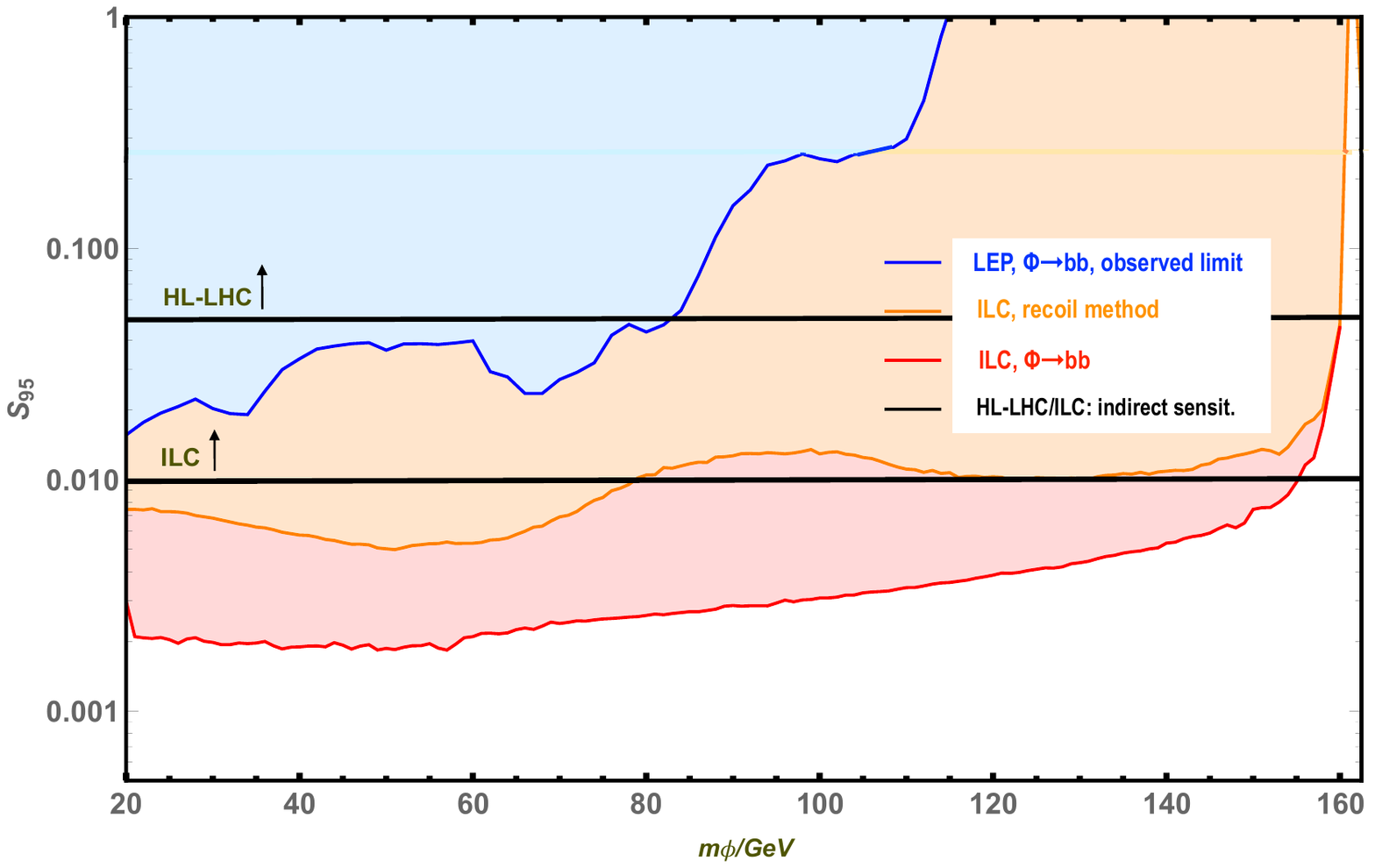}
\caption{\label{fig:lepgea} Sensitivity predictions for an ILC with a com energy of 250 \GeV~ from \cite{Drechsel:2018mgd}. See text for details.}
\end{center}
\end{figure}
\end{center}

\begin{center}
\begin{figure}[htb!]
\begin{center}
\includegraphics[width=0.45\textwidth]{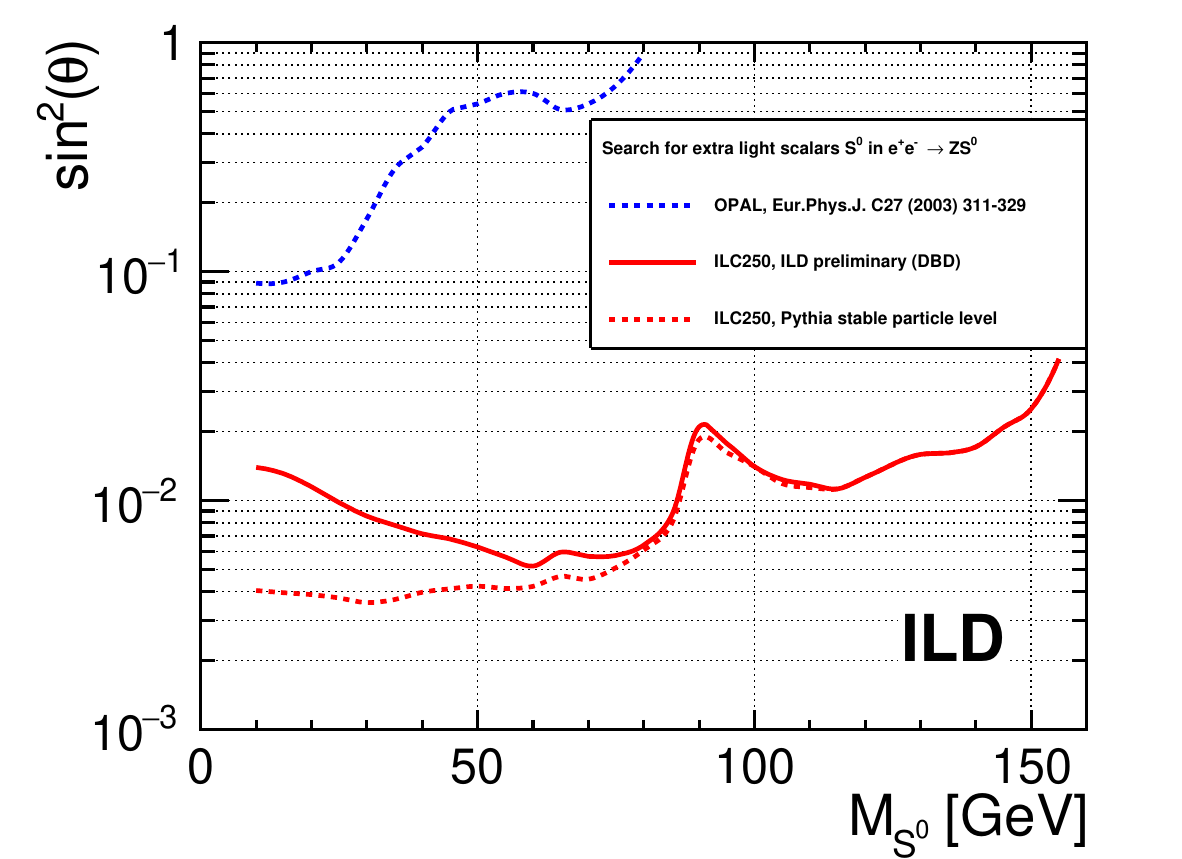}
\includegraphics[width=0.45\textwidth]{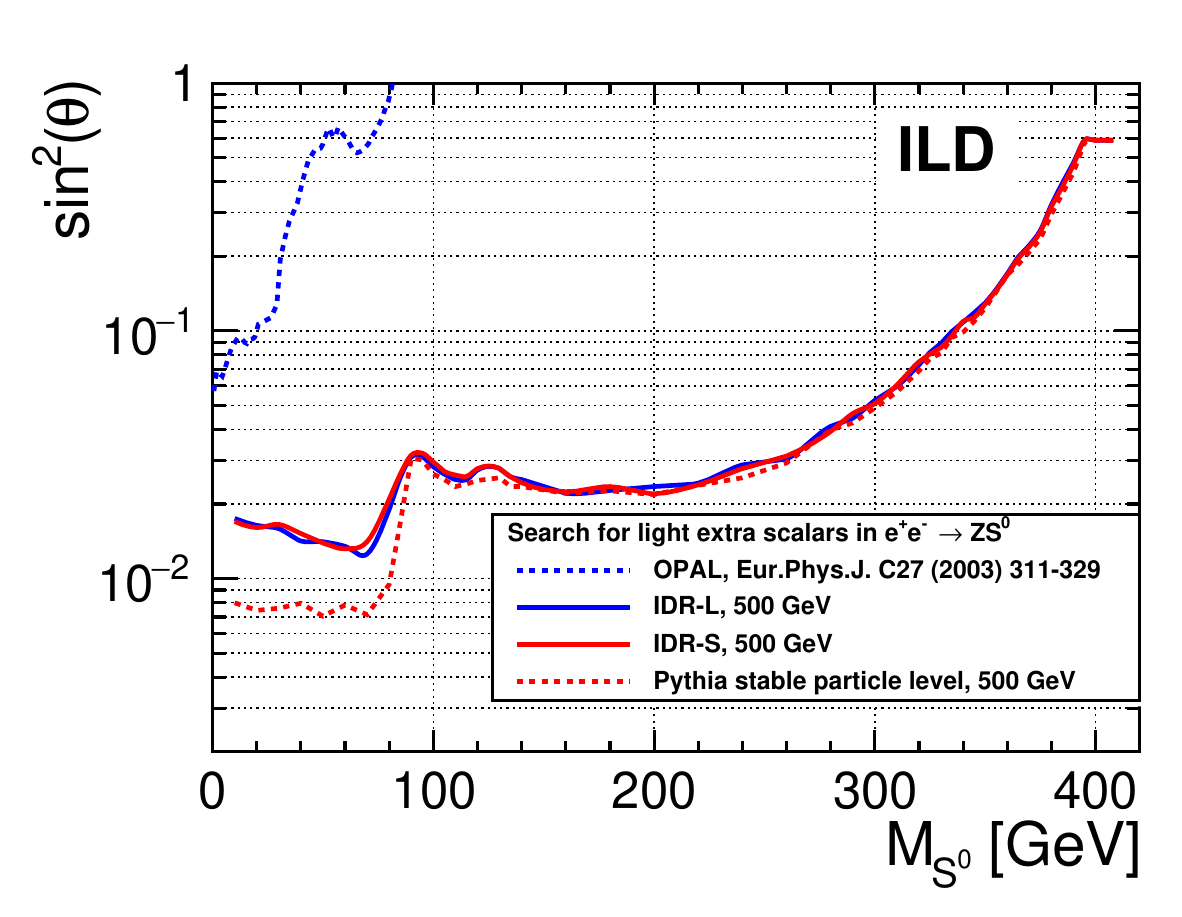}
\caption{\label{fig:jennyea} Upper bounds on the mixing angle for the model discussed in \cite{Wang:2020lkq}, in a comparison of different detector concepts and using the recoil method.}
\end{center}
\end{figure}
\end{center}

The production of lighter scalars in scalar strahlung has already been investigated in various works \cite{Drechsel:2018mgd,Wang:2020lkq}, where the latter focusses on the investigation of different detector concepts in an ILC environment. In principle two different analysis methods exist, which either use the pure $Z$ recoil ("recoil method") or take the light scalar decay into $b\,\bar{b}$ into account. In \cite{Drechsel:2018mgd}, the sensitivity of the ILC for low-mass scalars in $Z\,h$ was investigated. Figure \ref{fig:lepgea} shows the reach of these two methods at $95\,\%$ CL limit for agreement with a background only hypothesis, which can directly be translated into an upper bound on rescaling. The authors validate their method by reproducing LEP results \cite{LEPWorkingGroupforHiggsbosonsearches:2003ing,ALEPH:2006tnd} for these channels prior to applying their method to the ILC. 
Along similar lines, a more recent study that uses only the recoil method and compares different detector options has been presented in \cite{Wang:2020lkq}. The corresponding results are shown in figure \ref{fig:jennyea}. The authors perform their analysis in a model where the coupling of the new resonance is rescaled by a mixing angle $\sin\theta$ and the results can therefore be directly compared to the ones displayed in figure \ref{fig:lepgea}.

\section{Current enhancements and benchmark motivations}

\begin{center}
\begin{figure}[htb!]
\includegraphics[width=0.49\textwidth]{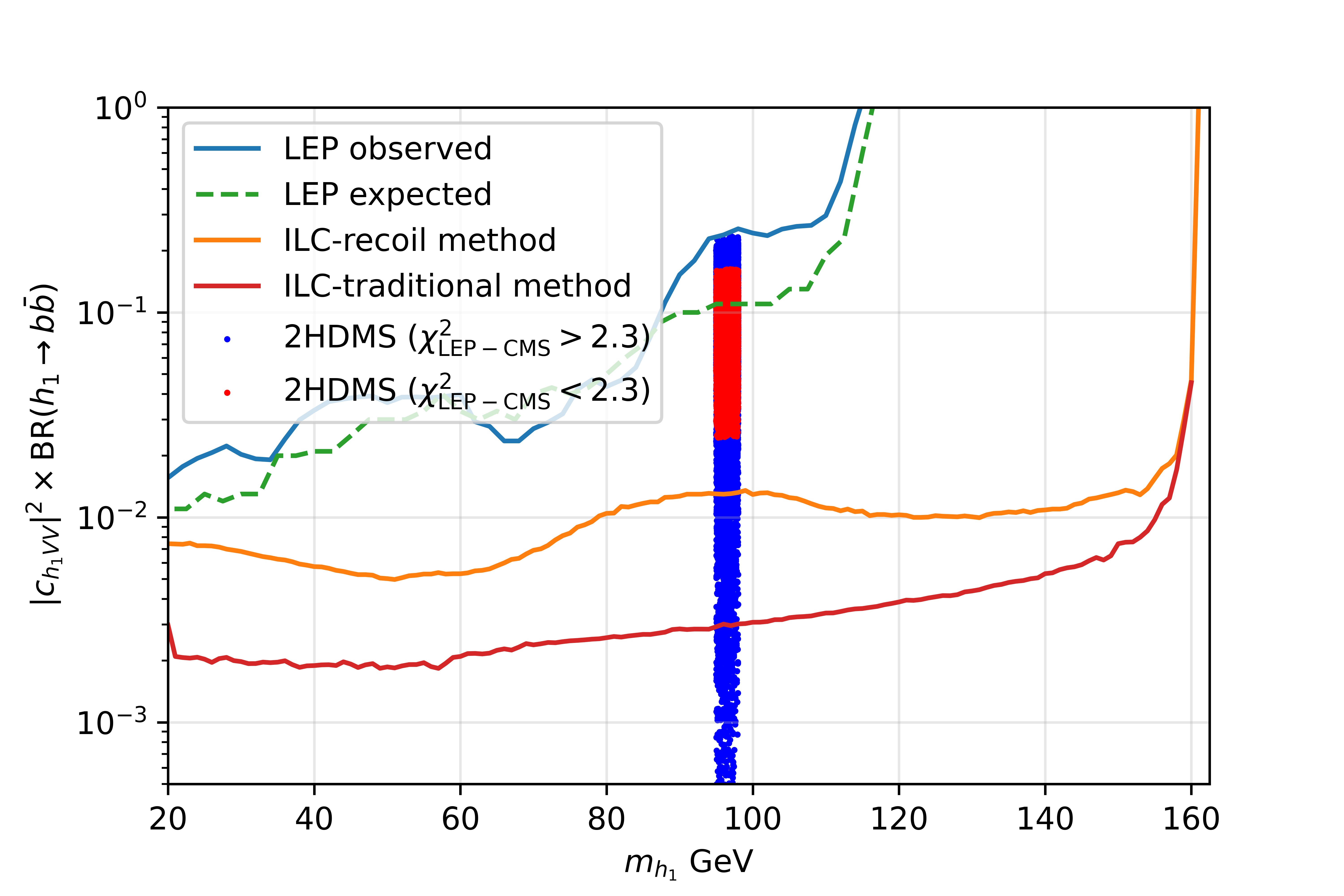}
\includegraphics[width=0.49\textwidth]{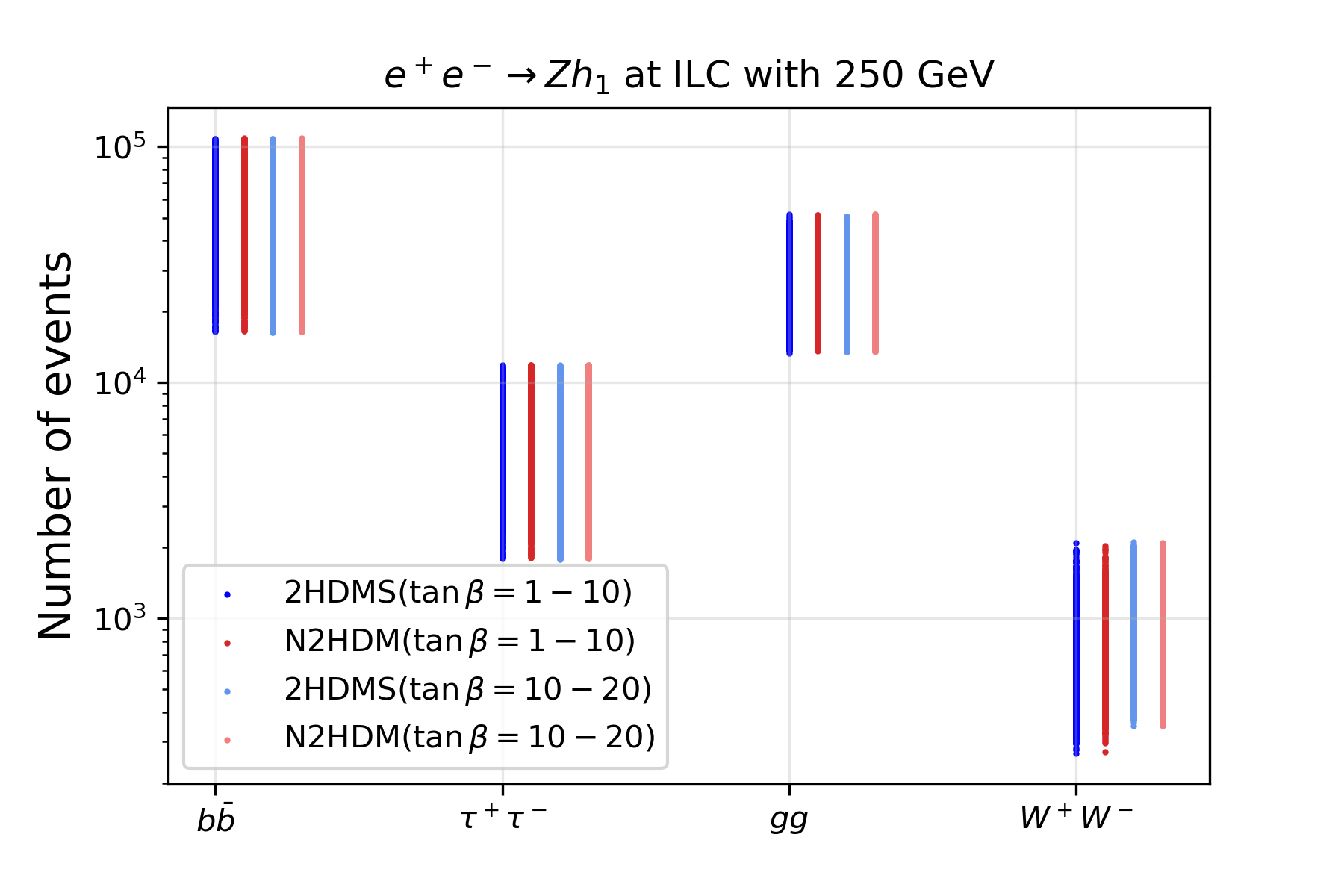}
\caption{\label{fig:svenea} {\sl Left:} Points in the 2HDMs that agree with both CMS and LEP excess and which can be probed at the ILC. {\sl Right:} predicted rates in the 2HDMS and N2HDM at 250 \GeV using full target luminosity.}
\end{figure}
\end{center}

Recently, there has been a lot of activity in the attempt to fit the possible low-mass excesses at LEP \cite{ALEPH:2006tnd} and CMS \cite{CMS:2018cyk} within new physics scenarios. I here mainly comment on the work presented in \cite{Heinemeyer:2021msz}, where several models are fitted to these excesses that contain singlet and doublet extensions of the SM scalar sector. The authors consider models with an additional doublet as well as a (complex) singlet, labelled N2HDM and 2HDMS, respectively. For both models, the authors investigate the possibility to fit the observed accesses. Out of the parameter space fitting these excesses, they then render rate predictions for a $250\,\GeV$ collider with a total luminosity of $\mathcal{L}\,=\,2\,\ab^{-1}$. The corresponding results are shown in figure \ref{fig:svenea}. In particular, it is of high interest that also other final states for the $h$ decay, as e.g. $\tau^+\tau^-,\,gg,$ or $W^+W^-$ can render sizeable rates. Related work concentrating on the N2HDM can be found in \cite{Biekotter:2022jyr}. See also \cite{Biekotter:2022abc,Benbrik:2022azi,Benbrik:2022dja,Li:2022etb,Azevedo:2023zkg,Escribano:2023hxj,Biekotter:2023oen,Belyaev:2023xnv,Ashanujjaman:2023etj} for other models that can in principle comply with an excess in this mass range. 
\section{Connection to electroweak phase transition}

Another important topic is the connection of models with extended scalar sectors with different scenarios of electroweak phase transitions. In particular, for scenarios where the second scalar is lighter than a SM like candidate, such states can be investigated in Higgs-strahlung and the associated decay of $h\,\rightarrow\,h_i\,h_i$. Due to the clean environment of a lepton collider, these processes can be investigated down to relatively low rates.

\begin{center}
\begin{figure}[htb!]
\includegraphics[width=0.45\textwidth]{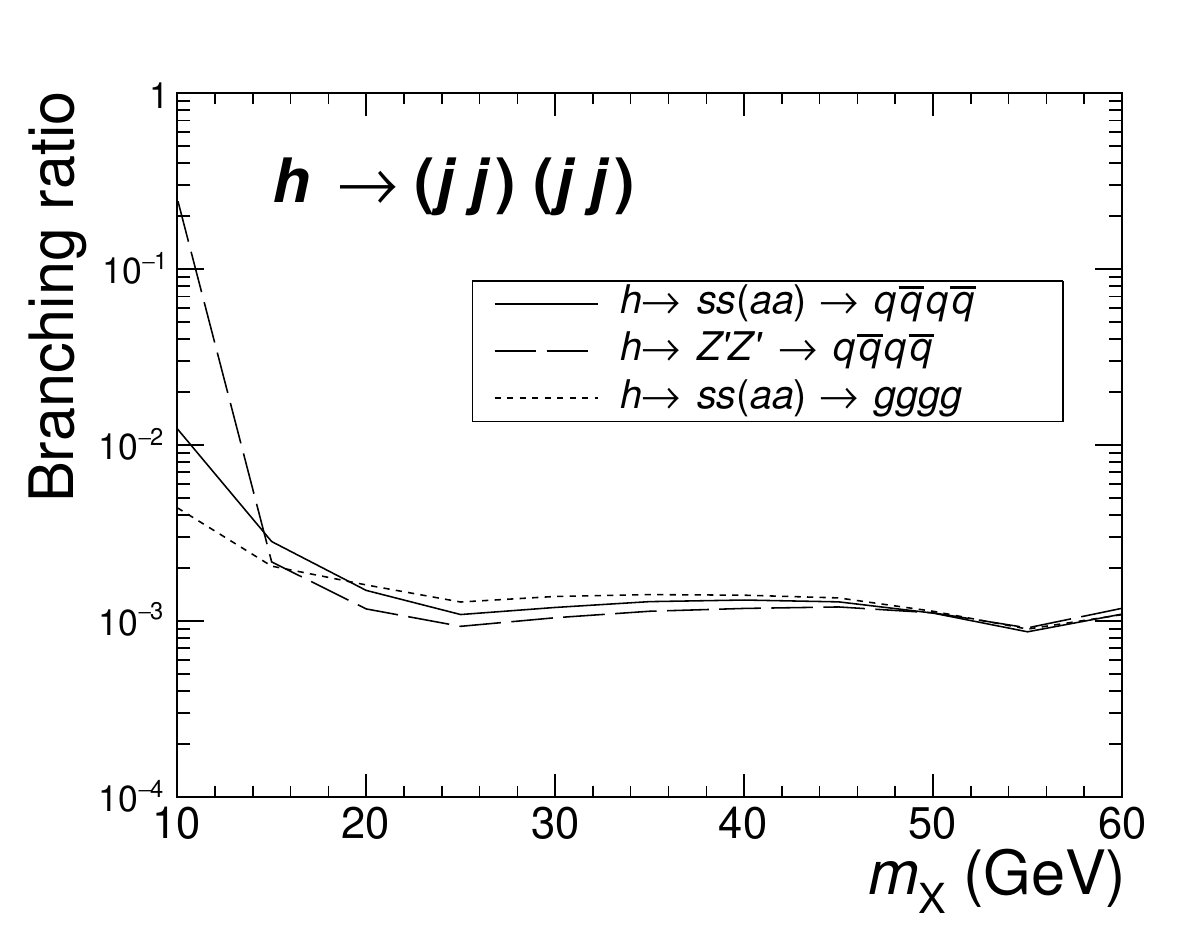}
\includegraphics[width=0.45\textwidth]{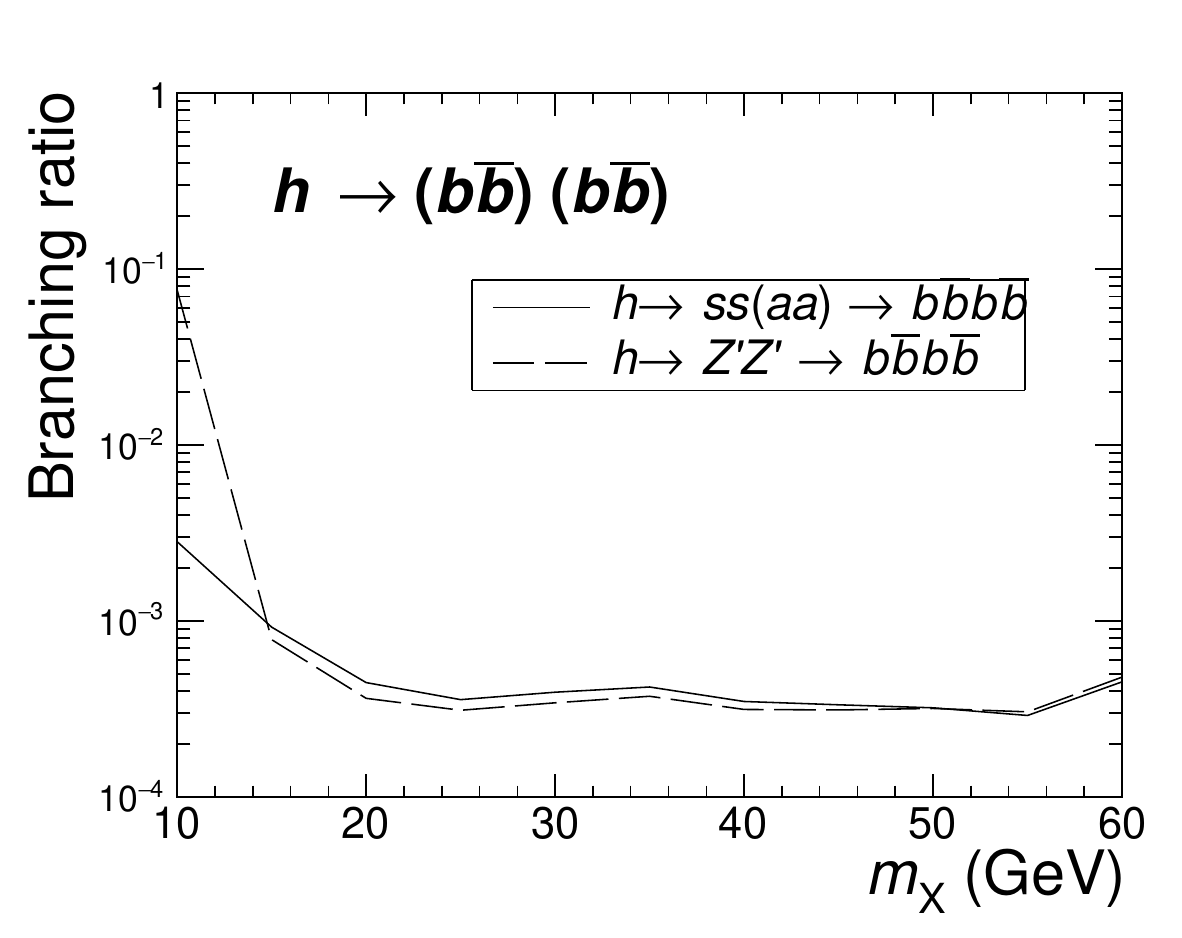}
\caption{\label{fig:discalar} 95 $\%$ confidence bounds on branching ratios for Higgs decay into a pair of lighter particles, for a com energy of 240~ \GeV and $\int\mathcal{L}\,=\,5\,\ab^{-1}$. Taken from \cite{Liu:2016zki}.}
\end{figure}
\end{center}

A standard reference for such processes is \cite{Liu:2016zki}, where the authors consider Higgs-strahlung at a 240 \GeV~ $e^+e^-$ collider, with the Higgs subsequently decays into two light scalar states corresponding to the above target signature. They give 95 $\%$ confidence level bounds for the branching ratios into the decay products of the two light scalars as a function of the light scalar masses for an integrated luminosity of $\int\mathcal{L}\,=\,5\,\ab^{-1}$ following a detailed study. We show their results for various channels in figure \ref{fig:discalar}.

Depending on the mass, model, and decay mode, branching ratios down to $\sim\,10^{-4}$ can be tested.

In simple singlet extensions it is possible to test regions in the models parameter space which can lead to a strong first-order electroweak phase transition. There has a been a lot if recent activity in this field recently; here we show results from \cite{Kozaczuk:2019pet}, where in addition several collider sensitivity projections are shown, including the bounds derived in \cite{Liu:2016zki}. It becomes clear that $e^+e^-$ Higgs factories would be an ideal environment to confirm or rule out such scenarios.

\begin{center}
\begin{figure}[htb!]
\begin{center}
\includegraphics[width=0.45\textwidth]{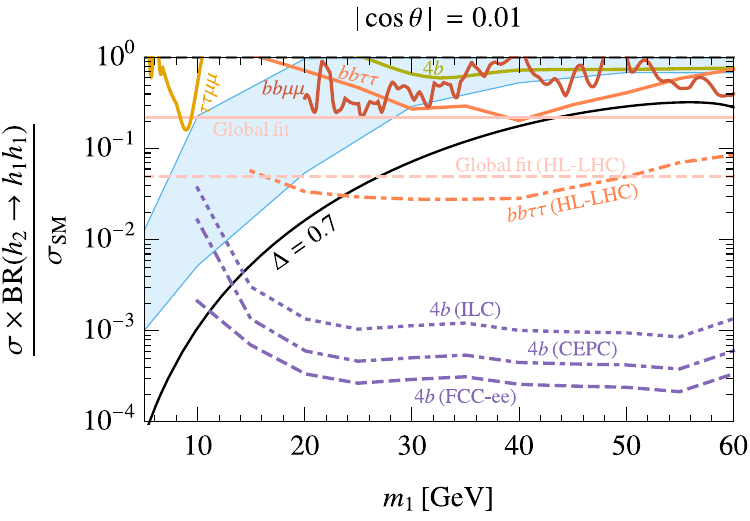}
\includegraphics[width=0.45\textwidth]{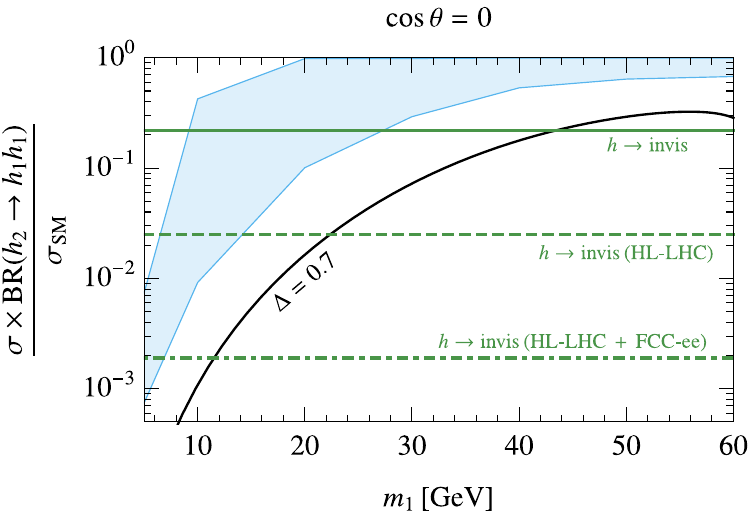}
\end{center}
\caption{Expected bounds on Higgs production via Higgs strahlung and subsequent decay into two light scalars, in the singlet extension scenario discussed in \cite{Kozaczuk:2019pet,Wang:2022dkz}. {\sl Left:} Taken from \cite{Kozaczuk:2019pet}. {\sl Right:} For $\cos\theta\,=\,0$ the constraints mainly stem from $h_{125}\,\rightarrow\,\text{invisble}$ searches. Depending on $m_1$ this scenario can be tested at current or future collideer experiments.}
\end{figure}
\end{center}

At the LHC, searches already exist for the above process. An overview can e.g. be found in \cite{Cepeda:2021rql}, where the authors present the status of current searches for 
\begin{\eqn}\label{eq:lims}
p\,p\,\rightarrow\,h_{125}\,\rightarrow\,h_i\,h_i\,\rightarrow\,X\,X\,Y\,Y
\end{\eqn}
which for such models can be read as a bound in 
\begin{\eqn*}
\sin^2\,\al\,\times\,\text{BR}_{h_{125}\,\rightarrow\,h_i\,h_i\,\rightarrow\,X\,X\,Y\,Y}.
\end{\eqn*}

\begin{center}
\begin{figure}
\begin{center}
\includegraphics[width=0.85\textwidth]{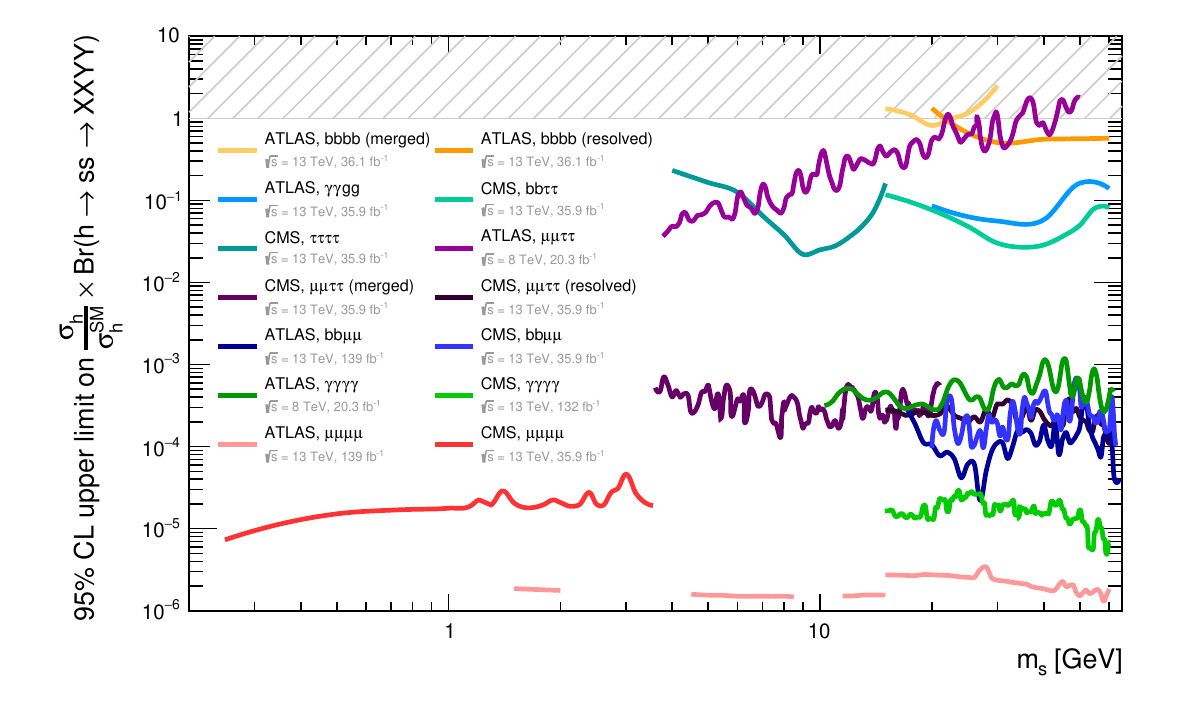}
\caption{\label{fig:current_stat} Limits on the process in eqn (\ref{eq:lims}), taken from \cite{Cepeda:2021rql}. This displays current constraints which can especially be easily reinterpreted in extended scalar sector models, in particular models where couplings are inherited via a simple mixing angle. In this figure, the lighter scalar is denoted by $s$, which corresponds to $h_i$ in the notation used in this manuscript.}
\end{center}
\end{figure}
\end{center}

The results are displayed in figure \ref{fig:current_stat}. Current bounds on the mixing angle for the 125 \GeV-like state are around $|\sin\,\al|\,\lesssim\,0.3$ (see e.g. \cite{Robens:2022cun}). This means that branching ratios $\text{BR}_{h_{125}\,\rightarrow\,h_i\,h_i\,\rightarrow\,X\,X\,Y\,Y}$ down to $\mathcal{O}\lb 10^{-5}\rb$ can be tested. In particular the $\mu\mu\mu\mu$ final states in the low mass region give interesting constraints on the $h_{125}\,\rightarrow\,h_i\,h_i$ branching ratio down to $\,\sim\,0.03$.
 See also \cite{Carena:2022yvx} for more recent work in this context.
\section{Parameter space for some sample models}

After investigating new physics signatures, we now turn to models that still allow for such low mass scalars. This is obviously only a brief overview, and more models might exist allong for low mass scalars accessible at Higgs factories; see e.g. \cite{Robens:2022zgk} for more details.

\begin{center}
\begin{figure}[htb!]
\begin{center}
\includegraphics[width=0.48\textwidth]{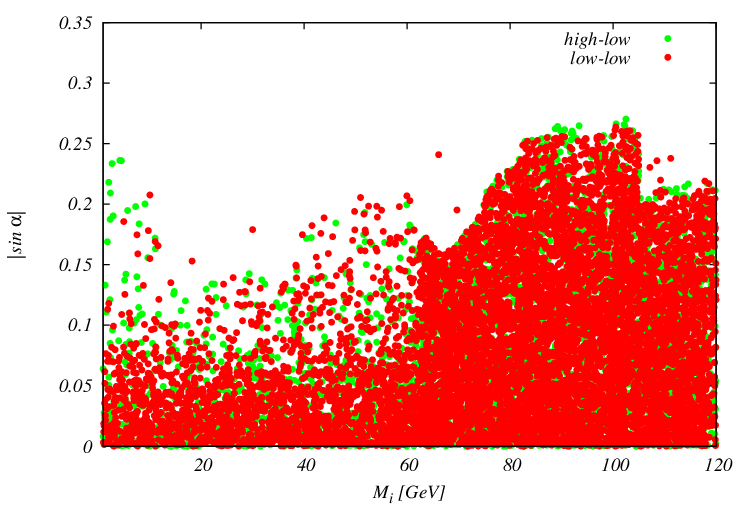}
\includegraphics[width=0.48\textwidth]{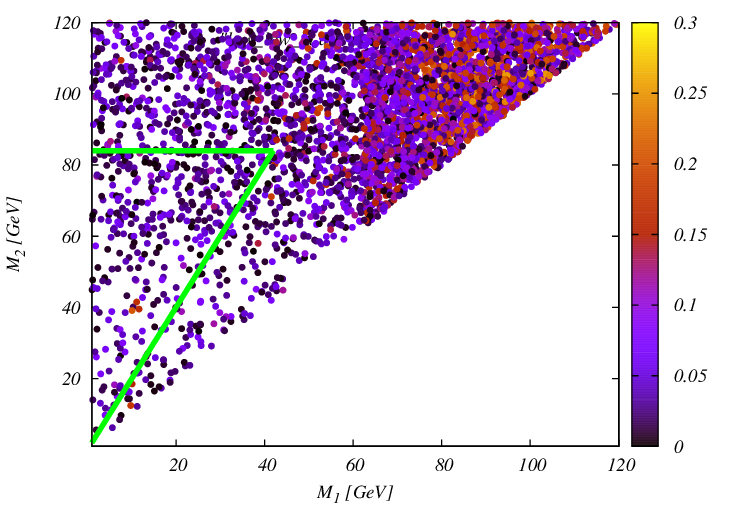}
\caption{\label{fig:trsm} Available parameter space in the TRSM, with one (high-low) or two (low-low) masses lighter than 125 \GeV. {\sl Left}: light scalar mass and mixing angle, with $\sin\al\,=\,0$ corresponding to complete decoupling. {\sl Right:} available parameter space in the $\lb m_{h_1},\,m_{h_2}\rb$ plane, with color coding denoting the rescaling parameter $\sin\al$ for the lighter scalar $h_1$. Within the green triangle, $h_{125}\,\rightarrow\,h_2 h_1\,\rightarrow\,h_1\,h_1\,h_1$ decays are kinematically allowed. Taken from \cite{Robens:2022zgk}.}
\end{center}
\end{figure}
\end{center}

The first model we discuss is a model that extends the scalar sector of the SM by two additional fields that transform as singlets under the electroweak gauge group \cite{Robens:2019kga,Robens:2022nnw}. This model contains three CP-even neutral scalars that relate the gauge and mass eigenstates $h_{1,2,3}$ via mixing, where one of the three scalars has to have properties complying with current measurements of the SM-like scalar, while the other two can have higher or lower masses. 
A detailed discussion including all constraints can be found in \cite{Robens:2019kga,Robens:2022nnw}, with recent updates on benchmark planes also presented in \cite{Robens:2023pax}. In figure \ref{fig:trsm}, two cases are shown where either one (high-low) or two (low-low) scalar masses are smaller than $125\,\GeV$. On the y-axis, the respective mixing angle is shown. Decoupling here corresponds to $\sin\al\,=\,0$.

\begin{center}
\begin{figure}[htb!]
\begin{center}
\includegraphics[width=0.48\textwidth]{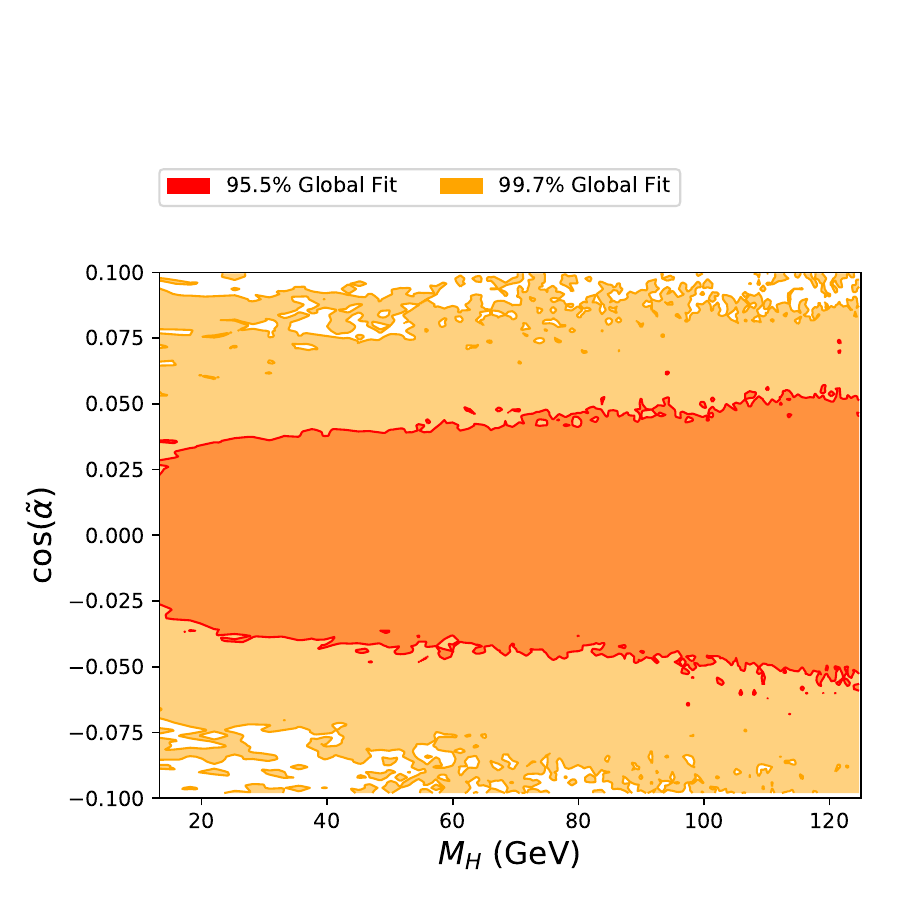} 
\includegraphics[width=0.48\textwidth]{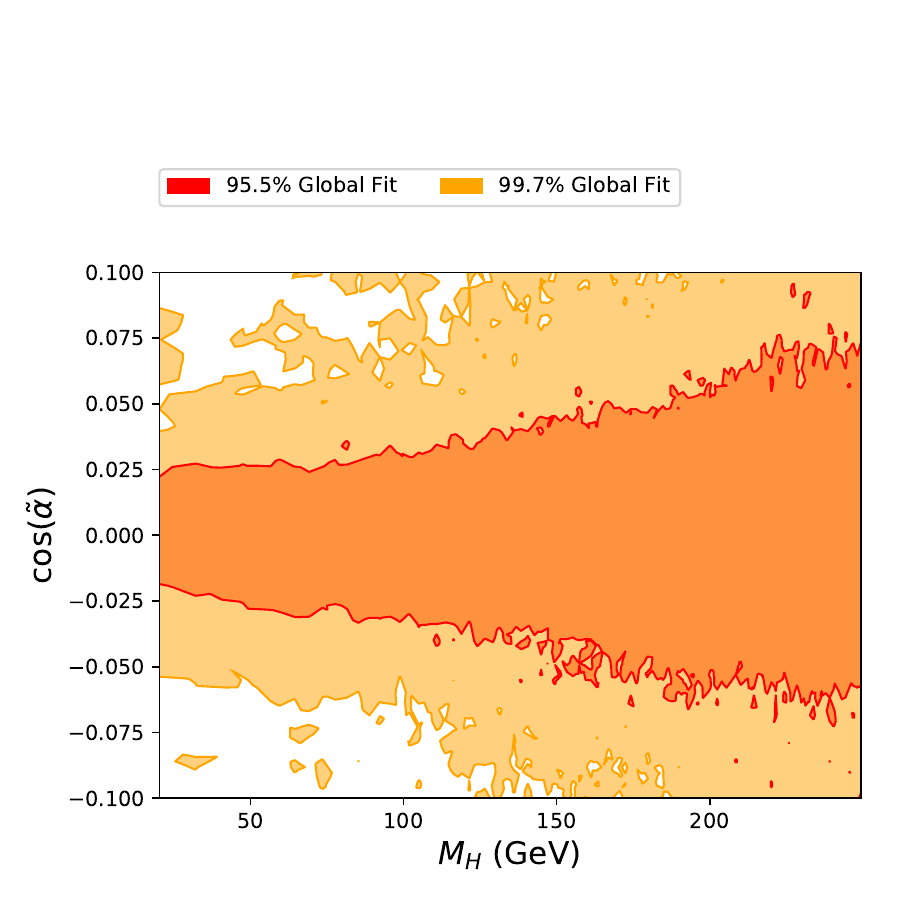}
\caption{\label{fig:victor} Mixing angle and masses of different additional scalars in the aligned 2HDM, from the scan presented in \cite{Eberhardt:2020dat}. For all additional scalars, regions exists where masses are $\lesssim\,125\,\GeV$, with absolute values of mixing angles such that $|\cos\lb\tilde{\al}\rb|\lesssim\,0.1$. Taken from \cite{Robens:2022zgk}.}
\end{center}
\end{figure}
\end{center}

Another option are e.g. two Higgs doublet models, where the SM scalar sector is augmented by a second doublet. In the so-called flavour-aligned scenario  \cite{Pich:2009sp,Pich:2010ic}, the authors perform a scan including bounds from theory, experimental searches and constraints, as e.g. electroweak observables, as well as B-physics. Here, the angle $\tilde{\al}$ parametrizes the rescaling with respect to the Standard Model couplings to gauge bosons, with $\cos\tilde{\al}\,=\,0$ designating the SM decoupling. The limits on the absolute value of the cosine of rescaling angle vary between 0.05 and 0.25 \cite{ATLAS-CONF-2021-053}. In figure \ref{fig:victor}, we show this angle vs the different scalar masses, reproduced from \cite{Eberhardt:2020dat} \footnote{I thank V. Miralles for providing these plots.}. 
We see that all regions for masses $\lesssim\,125\, \GeV$ can be populated, with absolute value of mixing angle ranges $|\cos\lb\tilde{\al}\rb|\lesssim\,0.1$. A more recent update of the above scan would of course be of high interest.
\begin{center}
\begin{figure}
\begin{center}
\includegraphics[width=0.95\textwidth]{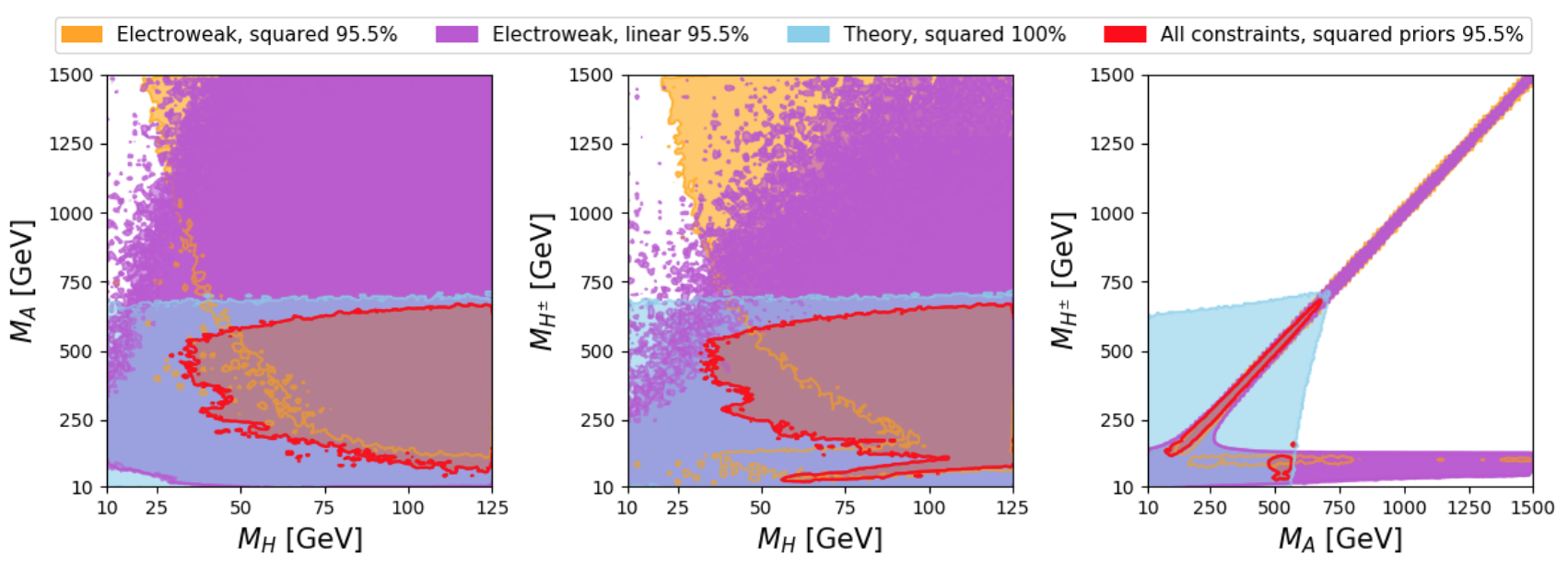}
\caption{Allowed regions in the 2HDM, from a scan presented in \cite{Eberhardt:2020dat}.}
\end{center}
\end{figure}
\end{center}

\begin{center}
\begin{figure}
\begin{center}
\includegraphics[width=0.6\textwidth]{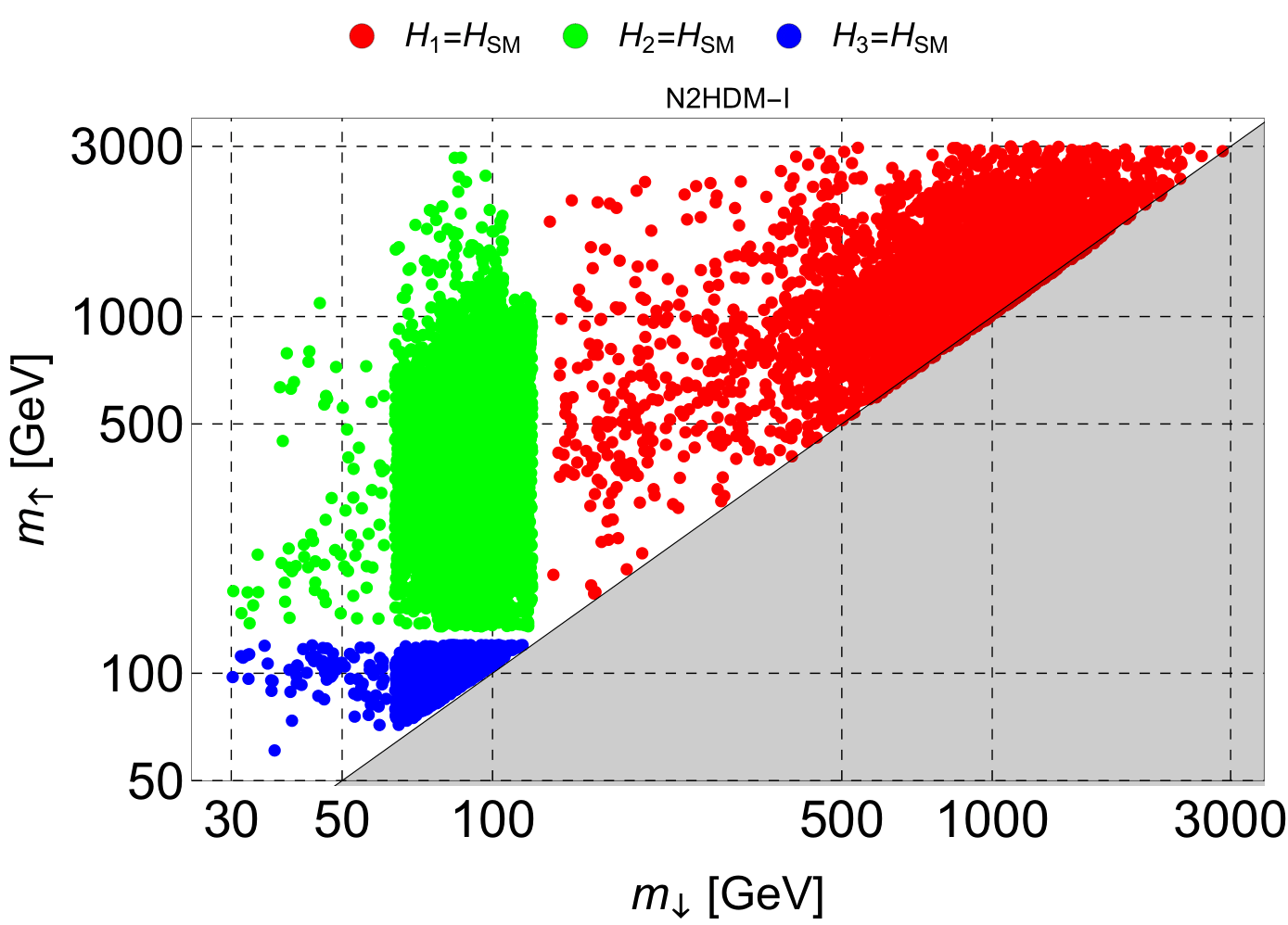}
\caption{\label{fig:n2hdm} Scan results in the N2HDM, taken from \cite{Abouabid:2021yvw}. There are regions in the models parameter space where either one or two of the additional scalars have masses $\lesssim\,125\,\GeV$.}
\end{center}
\end{figure}
\end{center}

On can also extend the scalar sector further, e.g. by adding an additional singlet in the gauge eigenbasis.  In \cite{Abouabid:2021yvw}, the authors consider a model where the SM scalar sector is extended by an additional doublet as well as a real singlet. The particle content of the model contains 3 CP even neutral scalar particles, out of which one, as before, needs to have the properties in compliance with LHC measurements of the 125 \GeV~ scalar. The authors perform an extensive scan and find regions in parameter space where either one or both of the additional scalars have masses below 125 \GeV. An example of the allowed parameter space is displayed in figure \ref{fig:n2hdm}. We see that in the CP-even sector there are regions within this model that still allow for low mass scalars.
Other models that allow for low mass scalars have e.g. been presented in \cite{deLima:2022yvn,Kuncinas:2023ycz}.
\section{Conclusions}
I very briefly discussed some aspects of searches for low mass scalars at Higgs factories, including models that allo for such low mass states, and provided references for further reading. In particular, novel studies exceeding the ones presented here are highly encouraged and could be included as an input for e.g. the next European Strategy update.

\begin{thebibliography}{10}

\bibitem{Robens:2022zgk}
Tania Robens.
\newblock {A Short Overview on Low Mass Scalars at Future Lepton Colliders}.
\newblock {\em Universe}, 8:286, 2022, 2205.09687.

\bibitem{Robens:2022uis}
Tania~Natalie Robens.
\newblock {An overview on low mass scalars at future lepton colliders}.
\newblock {\em PoS}, ICHEP2022:1041, 11 2022, 2211.10231.

\bibitem{EuropeanStrategyforParticlePhysicsPreparatoryGroup:2019qin}
Richard~Keith Ellis et~al.
\newblock {Physics Briefing Book}: {Input for the European Strategy for
  Particle Physics Update 2020}.
\newblock 10 2019, 1910.11775.

\bibitem{CERN-ESU-015}
{2020 Update of the European Strategy for Particle Physics (Brochure)}.
\newblock Technical report, Geneva, 2020.

\bibitem{OPAL:2002ifx}
G.~Abbiendi et~al.
\newblock {Decay mode independent searches for new scalar bosons with the OPAL
  detector at LEP}.
\newblock {\em Eur. Phys. J. C}, 27:311--329, 2003, hep-ex/0206022.

\bibitem{ALEPH:2006tnd}
S.~Schael et~al.
\newblock {Search for neutral MSSM Higgs bosons at LEP}.
\newblock {\em Eur. Phys. J. C}, 47:547--587, 2006, hep-ex/0602042.

\bibitem{Abramowicz:2016zbo}
H.~Abramowicz et~al.
\newblock {Higgs physics at the CLIC electron\textendash{}positron linear
  collider}.
\newblock {\em Eur. Phys. J. C}, 77(7):475, 2017, 1608.07538.

\bibitem{Alwall:2011uj}
Johan Alwall, Michel Herquet, Fabio Maltoni, Olivier Mattelaer, and Tim
  Stelzer.
\newblock {MadGraph 5 : Going Beyond}.
\newblock {\em JHEP}, 06:128, 2011, 1106.0522.

\bibitem{Drechsel:2018mgd}
P.~Drechsel, G.~Moortgat-Pick, and G.~Weiglein.
\newblock {Prospects for direct searches for light Higgs bosons at the ILC with
  250 GeV}.
\newblock {\em Eur. Phys. J. C}, 80(10):922, 2020, 1801.09662.

\bibitem{Wang:2020lkq}
Yan Wang, Mikael Berggren, and Jenny List.
\newblock {ILD Benchmark: Search for Extra Scalars Produced in Association with
  a $Z$ boson at $\sqrt{s}=500$ GeV}.
\newblock 5 2020, 2005.06265.

\bibitem{LEPWorkingGroupforHiggsbosonsearches:2003ing}
R.~Barate et~al.
\newblock {Search for the standard model Higgs boson at LEP}.
\newblock {\em Phys. Lett. B}, 565:61--75, 2003, hep-ex/0306033.

\bibitem{CMS:2018cyk}
Albert~M Sirunyan et~al.
\newblock {Search for a standard model-like Higgs boson in the mass range
  between 70 and 110 GeV in the diphoton final state in proton-proton
  collisions at $\sqrt{s}=$ 8 and 13 TeV}.
\newblock {\em Phys. Lett. B}, 793:320--347, 2019, 1811.08459.

\bibitem{Heinemeyer:2021msz}
S.~Heinemeyer, C.~Li, F.~Lika, G.~Moortgat-Pick, and S.~Paasch.
\newblock {Phenomenology of a 96~GeV Higgs boson in the 2HDM with an additional
  singlet}.
\newblock {\em Phys. Rev. D}, 106(7):075003, 2022, 2112.11958.

\bibitem{Biekotter:2022jyr}
Thomas Biek\"otter, Sven Heinemeyer, and Georg Weiglein.
\newblock {Mounting evidence for a 95 GeV Higgs boson}.
\newblock {\em JHEP}, 08:201, 2022, 2203.13180.

\bibitem{Biekotter:2022abc}
Thomas Biek\"otter, Sven Heinemeyer, and Georg Weiglein.
\newblock {Excesses in the low-mass Higgs-boson search and the ${W}$-boson mass
  measurement}.
\newblock {\em Eur. Phys. J. C}, 83(5):450, 2023, 2204.05975.

\bibitem{Benbrik:2022azi}
Rachid Benbrik, Mohammed Boukidi, Stefano Moretti, and Souad Semlali.
\newblock {Explaining the 96 GeV Di-photon anomaly in a generic 2HDM Type-III}.
\newblock {\em Phys. Lett. B}, 832:137245, 2022, 2204.07470.

\bibitem{Benbrik:2022dja}
Rachid Benbrik, Mohammed Boukidi, and Bouzid Manaut.
\newblock {$W$-mass and 96 GeV excess in type-III 2HDM}.
\newblock 4 2022, 2204.11755.

\bibitem{Li:2022etb}
Weichao Li, Jingya Zhu, Kun Wang, Shiquan Ma, Pengfu Tian, and Haoxue Qiao.
\newblock {A light Higgs boson in the NMSSM confronted with the CMS di-photon
  and di-tau excesses}.
\newblock 12 2022, 2212.11739.

\bibitem{Azevedo:2023zkg}
Duarte Azevedo, Thomas Biek\"otter, and P.~M. Ferreira.
\newblock {2HDM interpretations of the CMS diphoton excess at 95 GeV}.
\newblock 5 2023, 2305.19716.

\bibitem{Escribano:2023hxj}
Pablo Escribano, Victor~Martin Lozano, and Avelino Vicente.
\newblock {A Scotogenic explanation for the 95 GeV excesses}.
\newblock 6 2023, 2306.03735.

\bibitem{Biekotter:2023oen}
T.~Biek\"otter, S.~Heinemeyer, and G.~Weiglein.
\newblock {The 95.4 GeV di-photon excess at ATLAS and CMS}.
\newblock 6 2023, 2306.03889.

\bibitem{Belyaev:2023xnv}
Alexander Belyaev, Rachid Benbrik, Mohammed Boukidi, Manimala Chakraborti,
  Stefano Moretti, and Souad Semlali.
\newblock {Explanation of the Hints for a 95 GeV Higgs Boson within a 2-Higgs
  Doublet Model}.
\newblock 6 2023, 2306.09029.

\bibitem{Ashanujjaman:2023etj}
Saiyad Ashanujjaman, Sumit Banik, Guglielmo Coloretti, Andreas Crivellin, Bruce
  Mellado, and Anza-Tshilidzi Mulaudzi.
\newblock {$SU(2)_L$ triplet scalar as the origin of the 95 GeV excess?}
\newblock 6 2023, 2306.15722.

\bibitem{Liu:2016zki}
Zhen Liu, Lian-Tao Wang, and Hao Zhang.
\newblock {Exotic decays of the 125 GeV Higgs boson at future $e^+e^-$ lepton
  colliders}.
\newblock {\em Chin. Phys. C}, 41(6):063102, 2017, 1612.09284.

\bibitem{Kozaczuk:2019pet}
Jonathan Kozaczuk, Michael~J. Ramsey-Musolf, and Jessie Shelton.
\newblock {Exotic Higgs boson decays and the electroweak phase transition}.
\newblock {\em Phys. Rev. D}, 101(11):115035, 2020, 1911.10210.

\bibitem{Wang:2022dkz}
Zhen Wang, Xuliang Zhu, Elham~E. Khoda, Shih-Chieh Hsu, Nikolaos
  Konstantinidis, Ke~Li, Shu Li, Michael~J. Ramsey-Musolf, Yanda Wu, and
  Yuwen~E. Zhang.
\newblock {Study of Electroweak Phase Transition in Exotic Higgs Decays at the
  CEPC}.
\newblock In {\em {Snowmass 2021}}, 3 2022, 2203.10184.

\bibitem{Cepeda:2021rql}
Maria Cepeda, Stefania Gori, Verena~Martinez Outschoorn, and Jessie Shelton.
\newblock {Exotic Higgs Decays}.
\newblock 11 2021, 2111.12751.

\bibitem{Robens:2022cun}
Tania Robens.
\newblock {Constraining Extended Scalar Sectors at~Current and~Future
  Colliders\textemdash{}An Update}.
\newblock {\em Springer Proc. Phys.}, 292:141--152, 2023, 2209.15544.

\bibitem{Carena:2022yvx}
Marcela Carena, Jonathan Kozaczuk, Zhen Liu, Tong Ou, Michael~J. Ramsey-Musolf,
  Jessie Shelton, Yikun Wang, and Ke-Pan Xie.
\newblock {Probing the Electroweak Phase Transition with Exotic Higgs Decays}.
\newblock {\em LHEP}, 2023:432, 2023, 2203.08206.

\bibitem{Robens:2019kga}
Tania Robens, Tim Stefaniak, and Jonas Wittbrodt.
\newblock {Two-real-scalar-singlet extension of the SM: LHC phenomenology and
  benchmark scenarios}.
\newblock {\em Eur. Phys. J. C}, 80(2):151, 2020, 1908.08554.

\bibitem{Robens:2022nnw}
Tania Robens.
\newblock {Two-Real-Singlet-Model Benchmark Planes}.
\newblock {\em Symmetry}, 15(1):27, 2023, 2209.10996.

\bibitem{Robens:2023pax}
Tania Robens.
\newblock {Two-Real-Singlet Model Benchmark Planes -- A Moriond Update}.
\newblock In {\em {57th Rencontres de Moriond on QCD and High Energy
  Interactions}}, 5 2023, 2305.08595.

\bibitem{Pich:2009sp}
Antonio Pich and Paula Tuzon.
\newblock {Yukawa Alignment in the Two-Higgs-Doublet Model}.
\newblock {\em Phys. Rev. D}, 80:091702, 2009, 0908.1554.

\bibitem{Pich:2010ic}
Antonio Pich.
\newblock {Flavour constraints on multi-Higgs-doublet models: Yukawa
  alignment}.
\newblock {\em Nucl. Phys. B Proc. Suppl.}, 209:182--187, 2010, 1010.5217.

\bibitem{ATLAS-CONF-2021-053}
{Combined measurements of Higgs boson production and decay using up to $139$
  fb$^{-1}$ of proton-proton collision data at $\sqrt{s}= 13$ TeV collected
  with the ATLAS experiment}.
\newblock Technical report, CERN, Geneva, Nov 2021.
\newblock ATLAS-CONF-2021-053.

\bibitem{Eberhardt:2020dat}
Otto Eberhardt, Ana Pe\~nuelas Mart\'\i{}nez, and Antonio Pich.
\newblock {Global fits in the Aligned Two-Higgs-Doublet model}.
\newblock {\em JHEP}, 05:005, 2021, 2012.09200.

\bibitem{Abouabid:2021yvw}
Hamza Abouabid, Abdesslam Arhrib, Duarte Azevedo, Jaouad~El Falaki, Pedro.~M.
  Ferreira, Margarete M\"uhlleitner, and Rui Santos.
\newblock {Benchmarking di-Higgs production in various extended Higgs sector
  models}.
\newblock {\em JHEP}, 09:011, 2022, 2112.12515.

\bibitem{deLima:2022yvn}
Carlos~Henrique de~Lima and Heather~E. Logan.
\newblock {Unavoidable Higgs coupling deviations in the Z2-symmetric
  Georgi-Machacek model}.
\newblock {\em Phys. Rev. D}, 106(11):115020, 2022, 2209.08393.

\bibitem{Kuncinas:2023ycz}
A.~Kun\v{c}inas, O.~M. Ogreid, P.~Osland, and M.~N. Rebelo.
\newblock {Complex S$_{3}$-symmetric 3HDM}.
\newblock {\em JHEP}, 07:013, 2023, 2302.07210.

\end{thebibliography}


\begin{thebibliography}{10}



\end{thebibliography}

\end{document}